\documentclass[11pt]{article} 

\usepackage{textcomp, amssymb}  
\usepackage{anysize}            
\usepackage{amsfonts}
\usepackage{amsmath}
\usepackage{dsfont}
\usepackage{MnSymbol}
\usepackage{mathtools}
\usepackage{graphicx}
\usepackage{epsfig}
\usepackage{epstopdf}
\usepackage{tikz}
\usepackage{amsthm}
\usepackage{mathrsfs}
\usepackage{ifpdf}
\usepackage{thm-restate}
\usepackage{stmaryrd}
\usepackage{cleveref}
\usepackage{subcaption}
\usepackage{forest}
\usepackage{longtable}

\usetikzlibrary{calc,decorations.pathmorphing,shapes,graphs,positioning,automata,angles,cd,trees}

\newcounter{sarrow}

\newcounter{sarrow1}
\newcommand\xnrsquigarrow[1]{%
\stepcounter{sarrow1}%
\mathrel{\begin{tikzpicture}[baseline= {( $ (current bounding box.south) + (0,-0.5ex) $ )}]
\node[inner sep=.5ex] (\thesarrow) {$\scriptstyle #1$};
\path[draw,<-,decorate,
  decoration={zigzag,amplitude=0.7pt,segment length=1.2mm,pre=lineto,pre length=4pt}]
    (\thesarrow1.south east) -- (\thesarrow1.south west);
    $\slashedarrowfill@\relbar\relbar/$
    \end{tikzpicture}}%
}

\makeatletter
\def\slashedarrowfill@#1#2#3#4#5{%
  $\m@th\thickmuskip0mu\medmuskip\thickmuskip\thinmuskip\thickmuskip
   \relax#5#1\mkern-7mu%
   \cleaders\hbox{$#5\mkern-2mu#2\mkern-2mu$}\hfill
   \mathclap{#3}\mathclap{#2}%
   \cleaders\hbox{$#5\mkern-2mu#2\mkern-2mu$}\hfill
   \mkern-7mu#4$%
}
\def\rightslashedarrowfillb@{%
  \slashedarrowfill@\relbar\relbar/\rightarrow}
\newcommand\xnrightarrow[2][]{%
  \ext@arrow 0055{\rightslashedarrowfillb@}{#1}{#2}}

\def\rightslashedarrowfille@{%
  \slashedarrowfill@\relbar\relbar/\twoheadrightarrow}
\newcommand\xntworightarrow[2][]{%
  \ext@arrow 0055{\rightslashedarrowfille@}{#1}{#2}}

\def\rightslashedarrowfillg@{%
  \slashedarrowfill@\relbar\relbar{\raisebox{.12em}{}}\twoheadrightarrow}
\newcommand\xtworightarrow[2][]{%
  \ext@arrow 0055{\rightslashedarrowfillg@}{#1}{#2}}

\def\rightslashedarrowfillx@{%
  \slashedarrowfill@\Relbar\Relbar/\rightrightarrows}
\newcommand\xnTworightarrow[2][]{%
  \ext@arrow 0055{\rightslashedarrowfillx@}{#1}{#2}}

\def\rightslashedarrowfilly@{%
  \slashedarrowfill@\Relbar\Relbar{\raisebox{.12em}{}}\rightrightarrows}
\newcommand\xTworightarrow[2][]{%
  \ext@arrow 0055{\rightslashedarrowfilly@}{#1}{#2}}

\pgfdeclareshape{slash underlined}
{
  \inheritsavedanchors[from=rectangle] 
  \inheritanchorborder[from=rectangle]
  \inheritanchor[from=rectangle]{north}
  \inheritanchor[from=rectangle]{north west}
  \inheritanchor[from=rectangle]{north east}
  \inheritanchor[from=rectangle]{center}
  \inheritanchor[from=rectangle]{west}
  \inheritanchor[from=rectangle]{east}
  \inheritanchor[from=rectangle]{mid}
  \inheritanchor[from=rectangle]{mid west}
  \inheritanchor[from=rectangle]{mid east}
  \inheritanchor[from=rectangle]{base}
  \inheritanchor[from=rectangle]{base west}
  \inheritanchor[from=rectangle]{base east}
  \inheritanchor[from=rectangle]{south}
  \inheritanchor[from=rectangle]{south west}
  \inheritanchor[from=rectangle]{south east}
  \inheritanchorborder[from=rectangle]
  \foregroundpath{
    \southwest \pgf@xa=\pgf@x \pgf@ya=\pgf@y
    \northeast \pgf@xb=\pgf@x \pgf@yb=\pgf@y
    \pgf@xc=\pgf@xa
    \advance\pgf@xc by .5\pgf@xb
    \pgf@yc=\pgf@ya 
    \advance\pgf@xc by -1.3pt
    \advance\pgf@yc by -1.8pt
    \pgfpathmoveto{\pgfqpoint{\pgf@xc}{\pgf@yc}}
    \advance\pgf@xc by  2.6pt
    \advance\pgf@yc by  3.6pt
    \pgfpathlineto{\pgfqpoint{\pgf@xc}{\pgf@yc}}
    \pgfpathmoveto{\pgfqpoint{\pgf@xa}{\pgf@ya}}
    \pgfpathlineto{\pgfqpoint{\pgf@xb}{\pgf@ya}}
 }
}
\tikzset{nomorepostaction/.code=\let\tikz@postactions\pgfutil@empty}

\newcommand{\makecircled}[2][\mathord]{#1{\mathpalette\make@circled{#2}}}
\newcommand{\make@circled}[2]{%
  \begingroup\m@th
  \sbox\z@{$#1A$}%
  \sbox\tw@{%
    \raisebox{\depth}{%
      \vphantom{$#1A$}%
      \ooalign{%
        \hidewidth$#1\make@smaller{#1}{#2}$\hidewidth\cr
        $#1\bigcirc$\cr
      }%
    }%
  }%
  \resizebox{!}{\ht\z@}{\box\tw@}%
  \endgroup
}
\newcommand{\make@smaller}[2]{%
  \vcenter{\hbox{\scalebox{0.7}{$\m@th#1#2$}}}%
}
\makeatother

\newcommand*{\rmbrace}{|\mskip-4mu\}}
\newcommand*{\lmbrace}{\{\mskip-4mu|}
\newcommand*{\mset}[1]{\lmbrace#1\rmbrace}
\newcommand*{\rsbrace}{|\mskip-4mu\rangle}
\newcommand*{\lsbrace}{\langle\mskip-4mu|}
\newcommand*{\step}[1]{\lsbrace#1\rsbrace}

\newcommand{\pretesting}{\mathrel{\ooalign{\raise0.13ex\hbox{$\sqsubset$}\cr\hidewidth\raise-0.6ex\hbox{\scalebox{0.9}{$\sim$}}\hidewidth\cr}}}

\newcommand{\prepomset}{\lesssim_{\mathrm{P}}}

\newcommand{\prestep}{\lesssim_{\mathrm{S}}}

\newcommand{\prehp}{\lesssim_{\mathrm{HP}}}

\newcommand{\prehhp}{\lesssim_{\mathrm{HHP}}}

\newcommand{\prepomsets}[1]{\lesssim_{\mathrm{(P,#1)}}}

\newcommand{\presteps}[1]{\lesssim_{\mathrm{(S,#1)}}}

\newcommand{\prehps}[1]{\lesssim_{\mathrm{(HP,#1)}}}

\newcommand{\prehhps}[1]{\lesssim_{\mathrm{(HHP,#1)}}}

\newcommand{\preeq}{{\mathrel{\ooalign{\raise0.13ex\hbox{$=$}\cr\hidewidth\raise-0.6ex\hbox{\scalebox{0.9}{$\sim$}}\hidewidth\cr}}}}

\newcommand{\prepomseteq}{\preeq_{\mathrm{P}}}

\newcommand{\prestepeq}{\preeq_{\mathrm{S}}}

\newcommand{\prehpeq}{\preeq_{\mathrm{HP}}}

\newcommand{\prehhpeq}{\preeq_{\mathrm{HHP}}}

\newcommand{\preweakeq}{{\mathrel{\ooalign{\raise0.13ex\hbox{$=$}\cr\hidewidth\raise-0.6ex\hbox{\scalebox{0.9}{$\approx$}}\hidewidth\cr}}}}

\newcommand{\pretrace}{{\mathrel{\ooalign{\raise0.13ex\hbox{$\sim$}\cr\hidewidth\raise-0.6ex\hbox{\scalebox{0.9}{$<$}}\hidewidth\cr}}}}

\newtheorem{theorem}{Theorem}[section]
\newtheorem{definition}[theorem]{Definition}
\newtheorem{proposition}[theorem]{Proposition}
\newtheorem{lemma}[theorem]{Lemma}

\newtheorem{corollary}[theorem]{Corollary}

\begin{document}

\begin{titlepage}
\thispagestyle{empty}

\hrule
\begin{center}
{\bf\LARGE Finitary Truly Concurrent Bisimulations\\}

\vspace{0.5cm}
--- Yong Wang ---

\vspace{2cm}

\end{center}
\end{titlepage}

%
%

\newpage
\setcounter{page}{1}\pagenumbering{arabic}

        \section{Introduction}\label{intro}

Following the Scott-Strachey approach \cite{DS} to giving programming languages denotational semantics, many efforts have been discussed in the literature to give a process language a full abstract denotational semantics, i.e., the denotational model of a process language is in agreement with its operational intuition. Several bisimulation preorders have been developed, among those the famous one is prebisimulation \cite{Prebisimulation1} \cite{Prebisimulation2}, in which the divergence \cite{Divergence1} \cite{Divergence2} is considered.

To develop a full abstract denotational model of a process language based on prebisimulation preorder, its behavioural semantics has two problems: (1) Two processes related by a standard denotational interpretation afford the same finite observations. (2) Prebisimulation can make distinctions between the behaviours of two processes based on infinite observations. So, finitary part of prebisimulation is needed to obtain full abstract results.

There existed two main results on finitary bisimulation: the logical form \cite{DEB} and the behavioural form \cite{CFB}. Following the latter one \cite{CFB}, we give the definitions of truly concurrent prebisimulations and their finitary ones. 
\newpage\section{Truly Concurrent Bisimulations and Prebisimulations}\label{tcbp} 

\begin{definition}[Labelled poset]
A labelled poset is a tuple $\mathbf{u}=\langle S, \leq, \lambda\rangle$, where $S$ is the carrier set, $\leq$ is a partial order on $S$ and $\lambda$ is a labelling function $\lambda:S\rightarrow\Sigma$.

For a labelled poset $\mathbf{u}$, $S_{\mathbf{u}}$, $\leq_{\mathbf{u}}$ and $\lambda_{\mathbf{u}}$ denote the carrier, the partial order and the labelling of $\mathbf{u}$ respectively. The set of labelled posets is denoted $\mathsf{LP}$ and the empty labelled poset is $\mathbf{\epsilon}$.
\end{definition}

\begin{definition}[Poset morphism]
For posets $\langle A,\leq\rangle$ and $\langle A',\leq'\rangle$ and function $\varphi:A\rightarrow A'$, $f$ is called a poset morphism if for $a_0,a_1\in A$ with $a_0\leq a_1$, then $\varphi(a_0)\leq' \varphi(a_1)$ holds.
\end{definition}

\begin{definition}[Labelled poset isomorphism]
Let $\mathbf{u}=\langle S_1,\leq_1,\lambda_1\rangle$ and $\mathbf{v}=\langle S_2,\leq_2,\lambda_2\rangle$ be labelled posets. A labelled poset morphism $\varphi$ from $\mathbf{u}=\langle S_1,\leq_1,\lambda_1\rangle$ to $\mathbf{v}=\langle S_2,\leq_2,\lambda_2\rangle$ is a poset morphism from $\langle S_1,\leq_1\rangle$ and $\langle S_2,\leq_2\rangle$ with $\lambda_2\circ \varphi=\lambda_1$. Moreover, $\varphi$ is a labelled poset isomorphism if it is a bijection with $\varphi^{-1}$ is a poset isomorphism from $\langle S_2,\leq_2,\lambda_2\rangle$ to $\langle S_1,\leq_1,\lambda_1\rangle$. We say that $\mathbf{u}=\langle S_1,\leq_1,\lambda_1\rangle$ is isomorphic to $\mathbf{v}=\langle S_2,\leq_2,\lambda_2\rangle$ denoted $\langle S_1,\leq_1,\lambda_1\rangle\sim\langle S_2,\leq_2,\lambda_2\rangle$, if there exists a poset isomorphism $\varphi$ between $\langle S_1,\leq_1,\lambda_1\rangle$ and $\langle S_2,\leq_2,\lambda_2\rangle$.
\end{definition}

\begin{definition}[Multiset]
A multiset is a kind of set of objects which may be repetitive denoted $\mset{-}$, such that $\mset{a,b,b}$ is significantly distinguishable from $\mset{a,b}$.
\end{definition}

\begin{definition}[Pomset]
A partially ordered multiset, pomset, is a $\sim$-equivalence class of posets. The set of pomsets is denoted $\mathsf{Pom}$; the empty pomset is denoted $\mathbf{\epsilon}$ and the $\sim$-equivalence class of $\mathbf{\epsilon}$ is also denoted by $\epsilon$. If there does not exist partial orders between any two objects in a pomset, such pomset is called a step denoted $\step{-}$.The set of steps is denoted $\mathsf{Stp}$.
\end{definition}

\begin{definition}[Pomset labelled transition system]
A pomset labelled transition system (PLTS) is a quadruple $\langle\mathsf{Proc},\mathsf{Act},\{\xrightarrow{U}|U\in\mathsf{Pom}\},\mathsf{Pred}\rangle$, where:

\begin{enumerate}
  \item $\mathsf{Proc}$ is a set of processes, ranged over by $p,q,r,s$, which may be subscripted or superscripted.
  \item $\mathsf{Act}$ is a set of actions, ranged over by $a,b$, which may be subscripted or superscripted.
  \item $\mathsf{Pom}$ is the set of pomsets over $\mathsf{Act}$, ranged over by $U,V$, which may be subscripted or superscripted.
  \item $\xrightarrow{U}\subseteq\mathsf{Proc}\times\mathsf{Pom}\times\mathsf{Proc}$ is called a pomset transition for every $U\in\mathsf{Pom}$. We write $p\xrightarrow{U}q$ instead of $(p,q)\in\xrightarrow{U}$, and write $p\xnrightarrow{U}$ if $p\xrightarrow{U}q$ with no state $q$. Intuitively, $p\xrightarrow{U}q$ means that state $p$ can evolve into state $q$ by the execution of pomset $U$. We see that traditional single action transition $p\xrightarrow{a}q$ with $a\in\mathsf{Act}$ is a special case of pomset transition in which the pomset is primitive.
  \item For every $P\in\mathsf{Pred}$, we write $pP$ (resp. $p\neg P$) if state $p$ satisfies (resp. does not satisfy) predicate $P$. Intuitively, $pP$ means that predicate $P$ holds in state $p$. In $\mathsf{Pred}$, it includes a special kind of predicate called divergence, denoted $\uparrow$. $p\uparrow$ means that the information on $p$'s initial behaviours is incomplete. And we write $\downarrow$ for the convergence predicate, and $p\downarrow$ if and only if $p\nuparrow$, i.e., the information of $p$'s initial behaviours is completely specified.
\end{enumerate}

The binary pomset transitions $p\xrightarrow{U}p'$ and unary predicates $pP$ in a PLTS are called transitions.

Note that, by replacing $U\in\mathsf{Pom}$ by $a\in\mathsf{Act}$, we can get the definition of traditional labelled transition system (LTS). When $U\in\mathsf{Stp}$, we get the special case of a PLTS, called step labelled transition system (SLTS). And we use $\mathsf{Act}(p)\subseteq\mathsf{Act}$ to denote the set of actions occurring in $p$ and $\mathsf{Pom}(p)$ to denote the set of pomsets over $\mathsf{Act}(p)$.
\end{definition} 

For $n\geq 0$ and $\sigma=U_1\cdots U_n\in\mathsf{Pom}^*$, then $p\xrightarrow{\sigma}q$ if and only if there exist processes $p_0,\cdots,p_n$ such that $p=p_0\xrightarrow{U_1}p_1\xrightarrow{U_2}\cdots\xrightarrow{U_{n-1}}p_{n-1}\xrightarrow{U_n}p_n=q$. For $p\in\mathsf{Proc}$ and $U\in\mathsf{Pom}$, we define:

$$\mathsf{initials}(p)\triangleq\{U\in\mathsf{Pom}|\exists q:p\xrightarrow{U}q\}$$
$$\mathsf{sort}(p)\triangleq\{U\in\mathsf{Pom}|\exists\sigma\in\mathsf{Pom}^*,r,s\in\mathsf{Proc}:p\xrightarrow{\sigma}r\xrightarrow{U}s\}$$
$$\mathsf{derivatives}(p,U)\triangleq\{q|p\xrightarrow{U}q\}$$

A PLTS is sort-finite if and only if $\mathsf{sort}(p)$ is finite for every $p\in\mathsf{Proc}$.

\begin{definition}[Pomset synchronization tree]
The set of countably branching pomset synchronization trees over $\mathsf{Pom}$, denoted $\mathsf{ST}_{\infty}(\mathsf{Pom})$, is the set of infinitary terms generated by the following inductive definition:

$$\frac{\{U_i\in\mathsf{Pom},t_i\in\mathsf{ST}_{\infty}(\mathsf{Pom})\}_{i\in I}}{\sum_{i\in I}U_i:t_i[+\Omega]\in\mathsf{ST}_{\infty}(\mathsf{Pom})}$$

where $I$ is a countable index set, $\mathbb{O}=\sum_{i\in\emptyset}U_i:t_i$ which stands for the one-node synchronization tree, a representation of an inactive process, $\Omega=\sum_{i\in\emptyset}U_i:t_i+\Omega$ which stands that the behaviours of the synchronization tree are completely unspecified, $[+\Omega]$ means optional inclusion of $\Omega$ as a summand.
\end{definition} 

Let $\mathsf{ST}(\mathsf{Pom})$ denote the finite synchronization tree, i.e, the above set of terms is built using only finite summations. The set of synchronization trees $\mathsf{ST}_{\infty}(\mathsf{Pom})$ can be turned into a PLTS with divergence by stipulating that, for $t\in\mathsf{ST}_{\infty}(\mathsf{Pom})$:

\begin{itemize}
  \item $t\uparrow$ if and only if $\Omega$ is a summand of $t$.
  \item $t\xrightarrow{U_i}t_i$ if and only if $U_i:t_i$ is a summand of $t$.
\end{itemize} 

In the following, we give the definitions of truly concurrent bisimulations.

\begin{definition}[Configuration]
Let $\mathcal{P}=\langle\mathsf{Proc},\mathsf{Act},\{\xrightarrow{U}|U\in\mathsf{Pom}\},\mathsf{Pred}\rangle$ be a PLTS. A (finite) configuration in $\mathcal{P}$ is a (finite) consistent subset of actions $C\subseteq \mathcal{P}$, closed with respect to $\xrightarrow{U}$ (i.e. $\lceil C\rceil=C$). The set of finite configurations of $\mathcal{P}$ is denoted by $\mathcal{C}(\mathcal{P})$. For $p\in\mathsf{Proc}$, the corresponding configuration is denoted $C(p)$.
\end{definition}

\begin{definition}[Pomset, step bisimulation]\label{PSB}
Let $\langle\mathsf{Proc},\mathsf{Act},\{\xrightarrow{U}|U\in\mathsf{Pom}\},\mathsf{Pred}\rangle$ be a PLTS and $p,q\in\mathsf{Proc}$. A pomset bisimulation is a relation $R\subseteq\mathsf{Proc}\times\mathsf{Proc}$, such that:

\begin{enumerate}
  \item If $(p,q)\in R$, and $C(p)\xrightarrow{U}C(p')$ then $C(q)\xrightarrow{U}C(q')$, with $U\in \mathsf{Pom}$, and $(p',q')\in R$.
  \item If $(p,q)\in R$, and $C(q)\xrightarrow{U}C(q')$ then $C(p)\xrightarrow{U}C(p')$, with $U\in \mathsf{Pom}$, and $(p',q')\in R$.
\end{enumerate}

We say that $p$, $q$ are pomset bisimilar, written $p\sim_P q$, if there exists a pomset bisimulation $R$, such that $(p,q)\in R$. By replacing pomset transitions with steps, we can get the definition of step bisimulation. When $p$ and $q$ are step bisimilar, we write $p\sim_S q$.
\end{definition}

\begin{definition}[Posetal product]
Let $\langle\mathsf{Proc},\mathsf{Act},\{\xrightarrow{U}|U\in\mathsf{Pom}\},\mathsf{Pred}\rangle$ be a PLTS. Given two processes $p,q\in\mathsf{Proc}$, the posetal product of their configurations, denoted $\mathcal{C}(\mathcal{P})\overline{\times}\mathcal{C}(\mathcal{P})$, is defined as

$$\{(C(p),f,C(q))|C(p)\in\mathcal{C}(\mathcal{P}),C(q)\in\mathcal{C}(\mathcal{P}),f:C(p)\rightarrow C(q) \textrm{ isomorphism}\}$$

A subset $R\subseteq\mathcal{C}(\mathcal{P})\overline{\times}\mathcal{C}(\mathcal{P})$ is called a posetal relation. We say that $R$ is downward closed when for any $(C(p),f,C(q)),(C(p'),f',C(q'))\in \mathcal{C}(\mathcal{P})\overline{\times}\mathcal{C}(\mathcal{P})$, if $(C(p),f,C(q))\subseteq (C(p'),f',C(q'))$ pointwise and $(C(p'),f',C(q'))\in R$, then $(C(p),f,C(q))\in R$.

For $f:U\rightarrow U$, we define $f[a\mapsto a]:U\cup\{a\}\rightarrow U\cup\{a\}$, $z\in U\cup\{a\}$,(1)$f[a\mapsto a](z)=a$,if $z=a$;(2)$f[a\mapsto a](z)=f(z)$, otherwise. Where $U\in \mathsf{Pom}$, $a\in \mathsf{Act}$.
\end{definition}

\begin{definition}[(Hereditary) history-preserving bisimulation]\label{HHPB}
Let $\mathcal{P}=\langle\mathsf{Proc},\mathsf{Act},\{\xrightarrow{U}|U\in\mathsf{Pom}\},\mathsf{Pred}\rangle$ be a PLTS and $p,q\in\mathsf{Proc}$. A history-preserving (hp-) bisimulation is a posetal relation $R\subseteq\mathcal{C}(\mathcal{P})\overline{\times}\mathcal{C}(\mathcal{P})$ such that:

\begin{enumerate}
  \item If $(C(p),f,C(q))\in R$, and $C(p)\xrightarrow{a} C(p')$, then $C(q)\xrightarrow{a} C(q')$, with $(C(p'),f[a\mapsto a],C(q'))\in R$.
  \item If $(C(p),f,C(q))\in R$, and $C(q)\xrightarrow{a} C(q')$, then $C(p)\xrightarrow{a} C(p')$, with $(C(p'),f[a\mapsto a],C(q'))\in R$.
\end{enumerate}

$p,q$ are history-preserving (hp-)bisimilar and are written $p\sim_{HP}q$ if there exists an hp-bisimulation $R$ such that $(\emptyset,\emptyset,\emptyset)\in R$.

A hereditary history-preserving (hhp-)bisimulation is a downward closed hp-bisimulation. $p,q$ are hereditary history-preserving (hhp-)bisimilar and are written $p\sim_{HHP}q$.
\end{definition}

In the following, we give the definitions of truly concurrent prebisimulations.

\begin{definition}[Pomset, step prebisimulation]\label{PSPB}
Let $\langle\mathsf{Proc},\mathsf{Act},\{\xrightarrow{U}|U\in\mathsf{Pom}\},\mathsf{Pred}\rangle$ be a PLTS with divergence $\uparrow$ and $p,q\in\mathsf{Proc}$. Let $\mathsf{Rel}(\mathsf{Proc})$ denote the set of binary relations over $\mathsf{Proc}$. Define the functional $F:\mathsf{Rel}(\mathsf{Proc})\rightarrow\mathsf{Rel}(\mathsf{Proc})$, such that:

\begin{align*}
\begin{aligned}
F(R)=&\{(p,q)|\forall U\in\mathsf{Pom}\\
  &\bullet C(p)\xrightarrow{U}C(p')\Rightarrow C(q)\xrightarrow{U}C(q'), \textrm{ and }(p',q')\in R\\
  &\bullet p\downarrow\Rightarrow\bigg(q\downarrow\textrm{ and }\Big(C(q)\xrightarrow{U}C(q')\Rightarrow C(p)\xrightarrow{U}C(p'),\textrm{ and }(p',q')\in R\Big)\bigg)\}\\
\end{aligned}
\end{align*}

A relation $R$ is a pomset prebisimulation if and only if $R\subseteq F(R)$. We say that $p$, $q$ are pomset prebisimilar, written $p\prepomset q$, if there exists a pomset prebisimulation $R$, such that $(p,q)\in R$. By replacing pomset transitions with steps, we can get the definition of step prebisimulation. When $p$ and $q$ are step prebisimilar, we write $p\prestep q$.

The kernels of $\prepomset$ and $\prestep$ are denoted $\prepomseteq$ and $\prestepeq$.
\end{definition}

\begin{definition}[(Hereditary) history-preserving prebisimulation]\label{HHPPB}
Let $\langle\mathsf{Proc},\mathsf{Act},\{\xrightarrow{U}|U\in\mathsf{Pom}\},\mathsf{Pred}\rangle$ be a PLTS with divergence $\uparrow$ and $p,q\in\mathsf{Proc}$. Let $\mathsf{Rel}(\mathcal{C}(\mathcal{P})\overline{\times}\mathcal{C}(\mathcal{P}))$ denote the set of posetal relations over $\mathsf{Proc}$. Define the functional $F:\mathsf{Rel}(\mathcal{C}(\mathcal{P})\overline{\times}\mathcal{C}(\mathcal{P}))\rightarrow\mathsf{Rel}(\mathcal{C}(\mathcal{P})\overline{\times}\mathcal{C}(\mathcal{P}))$, such that:

\begin{align*}
\begin{aligned}
F(R)=&\{(C(p),f,C(q))|\forall a\in\mathsf{Act}\\
  &\bullet C(p)\xrightarrow{a}C(p')\Rightarrow C(q)\xrightarrow{a}C(q'), \textrm{ and }(C(p'),f[a\mapsto a],C(q'))\in R\\
  &\bullet p\downarrow\Rightarrow\bigg(q\downarrow\textrm{ and }\Big(C(q)\xrightarrow{a}C(q')\Rightarrow C(p)\xrightarrow{a}C(p'),\textrm{ and }(C(p'),f[a\mapsto a],C(q'))\in R\Big)\bigg)\}\\
\end{aligned}
\end{align*}

A relation $R$ is an hp-prebisimulation if and only if $R\subseteq F(R)$. We say that $p$, $q$ are hp-prebisimilar, written $p\prehp q$, if there exists an hp-prebisimulation $R$, such that $(\emptyset,\emptyset,\emptyset)\in R$. A hereditary history-preserving (hhp-)prebisimulation is a downward closed hp-prebisimulation. $p,q$ are hereditary history-preserving (hhp-)prebisimilar and are written $p\prehhp q$.

The kernels of $\prehp$ and $\prehhp$ are denoted $\prehpeq$ and $\prehhpeq$.
\end{definition}

\newpage\section{Finitary Truly Concurrent Prebisimulations}\label{ftcp}

By use of the functional $F$ in \cref{PSPB} and \cref{HHPPB}, we can obtain a pomset behavioural preorder which applies itself inductively as follows:

$$\prepomsets{0}\triangleq\mathsf{Proc}\times\mathsf{Proc}$$
$$\prepomsets{n+1}\triangleq F(\prepomsets{n})$$
$$\prepomsets{\omega}\triangleq\bigcap_{n\geq 0}\prepomsets{n}$$

Similarly, we can get the definitions of $\presteps{\omega}$, $\prehps{\omega}$ and $\prehhps{\omega}$.

\begin{definition}[Finitary pomset prebisimulation]
Let $\langle\mathsf{Proc},\mathsf{Act},\{\xrightarrow{U}|U\in\mathsf{Pom}\},\mathsf{Pred}\rangle$ be a PLTS and $p,q\in\mathsf{Proc}$. The finitary pomset prebisimulation $\prepomset^F$ over $\mathsf{Proc}$ is defined as follows: $p\prepomset^F q$ if and only if, for every $t\in\mathsf{ST}(\mathsf{Pom})$, $t\prepomset p$ implies $t\prepomset q$.
\end{definition} 

\begin{definition}[Finitary step prebisimulation]
Let $\langle\mathsf{Proc},\mathsf{Act},\{\xrightarrow{U}|U\in\mathsf{Pom}\},\mathsf{Pred}\rangle$ be a PLTS and $p,q\in\mathsf{Proc}$. The finitary step prebisimulation $\prestep^F$ over $\mathsf{Proc}$ is defined as follows: $p\prestep^F q$ if and only if, for every $t\in\mathsf{ST}(\mathsf{Pom})$, $t\prestep p$ implies $t\prestep q$.
\end{definition}

\begin{definition}[Finitary hp-prebisimulation]
Let $\langle\mathsf{Proc},\mathsf{Act},\{\xrightarrow{U}|U\in\mathsf{Pom}\},\mathsf{Pred}\rangle$ be a PLTS and $p,q\in\mathsf{Proc}$. The finitary hp-prebisimulation $\prehp^F$ over $\mathsf{Proc}$ is defined as follows: $p\prehp^F q$ if and only if, for every $t\in\mathsf{ST}(\mathsf{Pom})$, $t\prehp p$ implies $t\prehp q$.
\end{definition}

\begin{definition}[Finitary hhp-prebisimulation]
Let $\langle\mathsf{Proc},\mathsf{Act},\{\xrightarrow{U}|U\in\mathsf{Pom}\},\mathsf{Pred}\rangle$ be a PLTS and $p,q\in\mathsf{Proc}$. The finitary hhp-prebisimulation $\prehhp^F$ over $\mathsf{Proc}$ is defined as follows: $p\prehhp^F q$ if and only if, for every $t\in\mathsf{ST}(\mathsf{Pom})$, $t\prehhp p$ implies $t\prehhp q$.
\end{definition} 

They are obvious that $\prepomset~\subseteq~\prepomsets{\omega}~\subseteq~\prepomset^F$, $\prestep~\subseteq~\presteps{\omega}~\subseteq~\prestep^F$, $\prehp~\subseteq~\prehps{\omega}~\subseteq~\prehp^F$ and $\prehhp~\subseteq~\prehhps{\omega}~\subseteq~\prehhp^F$. Moreover, the inclusions are strict for infinitely branching PLTSs, and collapse to equalities for finitely branching PLTSs. 

Let $\langle\mathsf{Proc},\mathsf{Act},\{\xrightarrow{U}|U\in\mathsf{Pom}\},\mathsf{Pred}\rangle$ be a PLTS and $p,q\in\mathsf{Proc}$. For every $P\subseteq\mathsf{Pom}$, we define the following relations.

$\{\pretesting^{P}_{(P,n)}|n\geq 0\}$

$p\pretesting^{P}_{(P,0)}q\Leftrightarrow \mathrm{true}$
\begin{align*}
\begin{aligned}
&p\pretesting^{P}_{(P,n+1)}q\Leftrightarrow\\
  &\bullet \forall U\in P,C(p)\xrightarrow{U}C(p')\Rightarrow C(q)\xrightarrow{U}C(q'), \textrm{ and }p'\pretesting^{P}_{(P,n)}q'\\
  &\bullet \mathsf{initials(p)}\subseteq P\textrm{ and }p\downarrow\Rightarrow\bigg(\mathsf{initials(q)}\subseteq P\textrm,q\downarrow,\Big(\forall U\in P, C(q)\xrightarrow{U}C(q')\Rightarrow C(p)\xrightarrow{U}C(p'),\textrm{ and }p'\pretesting^{P}_{(P,n)}q'\Big)\bigg)\\
\end{aligned}
\end{align*}

Similarly, we can get the definition of $\{\pretesting^{P}_{(S,n)}|n\geq 0\}$, $\{\pretesting^{P}_{(HP,n)}|n\geq 0\}$ and $\{\pretesting^{P}_{(HHP,n)}|n\geq 0\}$.

\begin{proposition}
For every $n\geq 0$ and $P\subseteq\mathsf{Pom}$, the following statements hold:

\begin{enumerate}
  \item $\pretesting^{P}_{(P,n)}$, $\pretesting^{P}_{(S,n)}$, $\pretesting^{P}_{(HP,n)}$ and $\pretesting^{P}_{(HHP,n)}$ are all preorders.
  \item For $p,q\in\mathsf{Proc}$, 
  \begin{enumerate}
    \item $p\pretesting^{P}_{(P,n+1)}q\Rightarrow p\pretesting^{P}_{(P,n)}q$.
    \item $p\pretesting^{P}_{(S,n+1)}q\Rightarrow p\pretesting^{P}_{(S,n)}q$.
    \item $p\pretesting^{P}_{(HP,n+1)}q\Rightarrow p\pretesting^{P}_{(HP,n)}q$.
    \item $p\pretesting^{P}_{(HHP,n+1)}q\Rightarrow p\pretesting^{P}_{(HHP,n)}q$.
  \end{enumerate}
  \item Assume that $P\subseteq P'\subseteq\mathsf{Pom}$, then for $p,q\in\mathsf{Proc}$,
  \begin{enumerate}
    \item $p\pretesting^{P'}_{(P,n)}q\Rightarrow p\pretesting^{P}_{(P,n)}q$.
    \item $p\pretesting^{P'}_{(S,n)}q\Rightarrow p\pretesting^{P}_{(S,n)}q$.
    \item $p\pretesting^{P'}_{(HP,n)}q\Rightarrow p\pretesting^{P}_{(HP,n)}q$.
    \item $p\pretesting^{P'}_{(HHP,n)}q\Rightarrow p\pretesting^{P}_{(HHP,n)}q$.
  \end{enumerate}
\end{enumerate}
\end{proposition}

We define:

$$\pretesting^{P}_{(P,\omega)}\triangleq\bigcap_{n\geq 0}\pretesting^{P}_{(P,n)}$$
$$\pretesting^{P}_{(S,\omega)}\triangleq\bigcap_{n\geq 0}\pretesting^{P}_{(S,n)}$$
$$\pretesting^{P}_{(HP,\omega)}\triangleq\bigcap_{n\geq 0}\pretesting^{P}_{(HP,n)}$$
$$\pretesting^{P}_{(HHP,\omega)}\triangleq\bigcap_{n\geq 0}\pretesting^{P}_{(HHP,n)}$$
$$p\pretesting^{fin}_{(P,\omega)}q\triangleq\forall P\subseteq_{fin}\mathsf{Pom}.p\pretesting^{P}_{(P,\omega)}q$$
$$p\pretesting^{fin}_{(S,\omega)}q\triangleq\forall P\subseteq_{fin}\mathsf{Pom}.p\pretesting^{P}_{(S,\omega)}q$$
$$p\pretesting^{fin}_{(HP,\omega)}q\triangleq\forall P\subseteq_{fin}\mathsf{Pom}.p\pretesting^{P}_{(HP,\omega)}q$$
$$p\pretesting^{fin}_{(HHP,\omega)}q\triangleq\forall P\subseteq_{fin}\mathsf{Pom}.p\pretesting^{P}_{(HHP,\omega)}q$$

where $P\subseteq_{fin}\mathsf{Pom}$ means that $P$ is a finite subset of $\mathsf{Pom}$.

\begin{lemma}
For every $t\in\mathsf{ST}(\mathsf{Pom})$ and $p\in\mathsf{Proc}$, the following statements hold:

\begin{enumerate}
  \item $t\prepomset p\Leftrightarrow t\pretesting^{fin}_{(P,\omega)}p$.
  \item $t\prestep p\Leftrightarrow t\pretesting^{fin}_{(S,\omega)}p$.
  \item $t\prehp p\Leftrightarrow t\pretesting^{fin}_{(HP,\omega)}p$.
  \item $t\prehhp p\Leftrightarrow t\pretesting^{fin}_{(HHP,\omega)}p$.
\end{enumerate}
\end{lemma}

\begin{lemma}
The following statements hold:

\begin{enumerate}
  \item The preorder $\pretesting^{fin}_{(P,\omega)}$ is finitary.
  \item The preorder $\pretesting^{fin}_{(S,\omega)}$ is finitary.
  \item The preorder $\pretesting^{fin}_{(HP,\omega)}$ is finitary.
  \item The preorder $\pretesting^{fin}_{(HHP,\omega)}$ is finitary.
\end{enumerate}
\end{lemma}

\begin{theorem}
For $p,q\in\mathsf{Proc}$ in any transition system, the following statements hold:

\begin{enumerate}
  \item $p\prepomset^{F} q\Leftrightarrow p\pretesting^{fin}_{(P,\omega)}q$.
  \item $p\prestep^{F} q\Leftrightarrow p\pretesting^{fin}_{(S,\omega)}q$.
  \item $p\prehp^{F} q\Leftrightarrow p\pretesting^{fin}_{(HP,\omega)}q$.
  \item $p\prehhp^{F} q\Leftrightarrow p\pretesting^{fin}_{(HHP,\omega)}q$.
\end{enumerate}
\end{theorem}

\begin{corollary}
For $p,q\in\mathsf{Proc}$ in any sort-finite transition system, the following statements hold:

\begin{enumerate}
  \item $p\prepomset^{F} q\Leftrightarrow p\prepomsets{\omega}q$.
  \item $p\prestep^{F} q\Leftrightarrow p\presteps{\omega}q$.
  \item $p\prehp^{F} q\Leftrightarrow p\prehps{\omega}q$.
  \item $p\prehhp^{F} q\Leftrightarrow p\prehhps{\omega}q$.
\end{enumerate}
\end{corollary} 
\bibliographystyle{elsarticle-num}
\newpage\bibliography{Refs-CFMAI}

\begin{thebibliography}{1}
\expandafter\ifx\csname url\endcsname\relax
  \def\url#1{\texttt{#1}}\fi
\expandafter\ifx\csname urlprefix\endcsname\relax\def\urlprefix{URL }\fi
\expandafter\ifx\csname href\endcsname\relax
  \def\href#1#2{#2} \def\path#1{#1}\fi

\bibitem{DS}
D.~Scott, C.~Strachey, Towards a mathematical semantics for computer languages,
  in: J.~Fox (Ed.), Proceedings of the Symposium on Computers and Automata,
  Vol.~21 of MRI Symposium Proceedings, Polytechnic Press, Brooklyn, New York,
  1971, pp. 19--46.

\bibitem{Prebisimulation1}
M.~Hennessy, G.~D. Plotkin, A term model for {CCS}, in: MFCS 1980, Vol.~88 of
  LNCS, Springer, 1980, pp. 261--274.
\newblock \href {https://doi.org/10.1007/BFb0022510}
  {\path{doi:10.1007/BFb0022510}}.

\bibitem{Prebisimulation2}
M.~Hennessy, A term model for synchronous processes, Inf. Control. 51~(1)
  (1981) 58--75.
\newblock \href {https://doi.org/10.1016/S0019-9958(81)90044-5}
  {\path{doi:10.1016/S0019-9958(81)90044-5}}.

\bibitem{Divergence1}
D.~Walker, Bisimulation and divergence, Information and Computation 85~(2)
  (1990) 202--241.
\newblock \href {https://doi.org/10.1016/0890-5401(90)90013-7}
  {\path{doi:10.1016/0890-5401(90)90013-7}}.

\bibitem{Divergence2}
L.~Aceto, M.~Hennessy, Termination, deadlock, and divergence, J. ACM 39~(1)
  (1992) 147--187.
\newblock \href {https://doi.org/10.1145/147508.147527}
  {\path{doi:10.1145/147508.147527}}.

\bibitem{DEB}
S.~Abramsky, A domain equation for bisimulation, Inf. Comput. 92~(2) (1991)
  161--218.
\newblock \href {https://doi.org/10.1006/inco.1991.9999}
  {\path{doi:10.1006/inco.1991.9999}}.

\bibitem{CFB}
L.~Aceto, A.~Ing\'{o}lfsd\'{o}ttir, A characterization of finitary
  bisimulation, Information Processing Letters 64~(3) (1997) 127--134.

\end{thebibliography}

\end{document}